\documentclass[usenatbib]{mn2e}
\usepackage[dvips]{color}
\usepackage{graphicx}
\usepackage{lscape}
\title{Completing the puzzle of the 2004--2005 outburst in V\,0332+53: the brightening phase included}
\author[Tsygankov, Lutovinov, \&\ Serber]{S.\,S.\,Tsygankov$^{1,2}$\thanks{E-mail:
sst@mpa-garching.mpg.de},
A.\,A.\,Lutovinov$^{2}$, and A.\,V.\,Serber$^{3}$\\
$^{1}$MPI for Astrophysics, Karl-Schwarzschild str. 1, Garching, 85741, Germany\\
$^{2}$Space Research Institute of the Russian Academy of Sciences, Profsoyuznaya str. 84/32, Moscow
  117997, Russia\\
$^{3}$Institute of Applied Physics of the Russian Academy of Sciences,
46 Ulyanov st., 603950 Nizhny Novgorod, Russia}
\begin{document}

\date{Accepted .... Received ...}

\pagerange{\pageref{firstpage}--\pageref{lastpage}} \pubyear{2009}

\maketitle

\label{firstpage}

\begin{abstract}

%\textcolor{green}

{Analysis of the data obtained with the RXTE observatory during a powerful
outburst of the X-ray pulsar V\,0332+53 in 2004--2005 is presented.
Observational data covering the outburst brightening phase are analysed in
detail for the first time. A comparison of source parameters and their
evolution during the brightening and fading phases shows no evidence for any
hysteresis behaviour. It is found that the dependences of the energy of the
cyclotron absorption line on the luminosity during the brightening and
fading phases are almost identical. The complete data sequence including the
outburst brightening and fading phases makes it possible to impose the more
stringent constraints on the magnetic field in the source. The pulse profile
and pulsed fraction are studied as functions of the luminosity and
photon energy.}  \end{abstract}

\begin{keywords}
X-ray:binaries -- (stars:)pulsars:individual -- V\,0332+53
\end{keywords}

\section{Introduction}

It is conventionally believed that the structure of accretion environment of
a strongly magnetized neutron star in an X-ray pulsar, as well as its
emission, are actually governed by a few key parameters such as the
neutron-star spin period and magnetic field, the binary orbital period and
separation of the companions, and of course the accretion rate
$\dot{M}$. Sporadic outbursts occurring in some X-ray pulsars when the
bolometric X-ray luminosity $L_\mathrm{X}$ increases due to an enhanced
accretion rate provide for the possibility of probing the observable
response of such a complex system as an accreting X-ray pulsar to varying
$\dot{M}$ and getting more insight into the accretion onto a strongly
magnetized neutron star and the accompanying high-energy emission.

In particular, it is known since GINGA observations of the X-ray pulsar
4U\,0115+63 in 1990--1991 \citep{mih98} that the energies of cyclotron
absorption features observed in X-ray pulsar spectra can vary with their
luminosities. Such variations are conventionally interpreted in terms of the
luminosity dependence of the height of the line-forming region above the
neutron-star surface \citep{bs76}. The correspondent negative correlation
between the cyclotron-line energy and the luminosity was observed to
date for several bright X-ray pulsars (e.g. \citet{mih98,tsy06}). Note
that the opposite case of the positive correlation was also registered at
least for one or two X-ray pulsars (e.g. see \citet{sta07} for Her X-1 and
\citet{lab05,fil05} for GX301-2).

It should be emphasized that, to date, cyclotron-line parameters for
all sources were mostly studied during fading phases of outbursts
because of time lags between the beginning of an outburst and the
start of observations. For example, it was shown by \citet{tsy06}
(hereafter referred to as Paper~I) that the energy of the cyclotron
line observed in spectra of the X-ray pulsar V\,0332+53 increases
almost linearly with decreasing luminosity during the fading
phase of an outburst in 2004--2005.

This paper presents a comprehensive analysis of temporal and
spectral parameters of V\,0332+53 over a nearly complete powerful
outburst of 2004--2005, including the brightening phase studied in
detail for the first time.

\section{Observations and Data Analysis}

This work employs the RXTE data \citep{br93} acquired during the observations
with the following IDs: 90014-XX-XX-XX, 90089-XX-XX-XX
and 90427-XX-XX-XX. Note that Paper~I was based only on the ID
90427-XX-XX-XX data covering the fading phase of the outburst. Since the
aims of the present work and Paper~I are slightly different, these
data were reanalyzed to ensure the uniformity with the IDs
90089-XX-XX-XX and 90014-XX-XX-XX corresponding to the brightening
and fading phases of the outburst, respectively.

\begin{figure}
\includegraphics[width=\columnwidth,clip]{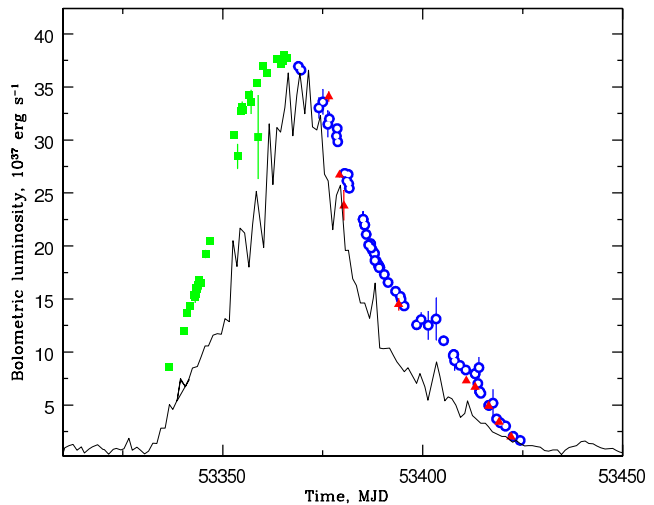}
\caption{Temporal dependence of the V\,0332+53 luminosity during the
outburst of 2004--2005. Filled green squares and open blue circles
correspond to the brightening and fading phases, respectively. The triangles
are INTEGRAL results. The solid line is the rescaled ASM light curve. The
error bars here and in the entire paper correspond to
$\pm\!1\sigma$.}\label{lcurve} \end{figure}

In general, the data-analysis procedure was identical to that used
in Paper~I. The luminosity was calculated assuming a source distance
of 7~kpc \citep{neg99}. Note that the correct account of the dead
time is crucial since the source count rate reached very high
values. Using improved software, we recalculated the luminosities
for the observations in the ID~90427 set and found that the
luminosities for a few brightest epochs of this set are to be
lowered by about 15--20\% in comparison with the values published in
Paper~I. Note, these corrections are not too large to affect
the conclusions of Paper~I. The time dependence of the bolometric
luminosity in the photon energy range $3\!-\!100$~keV during
the outburst of 2004--2005 is shown in Fig.~\ref{lcurve}. All the
data are naturally grouped into the brightening (filled green
squares) and fading phases (open blue circles) of the outburst. For
comparison, the result of INTEGRAL observations are also plotted by
the triangles in this figure (see Paper~I). The solid line is the
rescaled light curve obtained with the ASM monitor in the photon
energy range $2\!-\!12$\,keV. Further data reduction for both PCA
and HEXTE spectrometers was done using standard tools of
FTOOLS/LHEASOFT 6.3.2 package.

Since the source was bright and the outburst was nearly completely covered
with the RXTE observations (see Fig.~\ref{lcurve}), it is possible to obtain
the high-quality broadband energy spectra of the source at different stages
of the outburst and study the evolution of source parameters with the
luminosity and the outburst phase. It was noted in Paper~I that the best-fit
cutoff energy $E_\mathrm{cut}\!\simeq\!5\!-\!6$\,keV for the XSPEC {\em
powerlaw$\times$highecut} model \citep{wh83}, which is usually used for
approximation of X-ray pulsars spectra, is close to the lower limit
($\sim$3~keV) of the PCA energy range, so that the photon index cannot be
accurately retrieved from such a continuum model. Hence, the continuum
spectrum of the source was fitted with the simpler {\em cutoffpl} model
which describes the source spectrum similarly well, but has fewer parameters
and is free of the above limitations.  The continuum spectrum was modified
by one or few cyclotron-harmonic line features depending on the source
luminosity. A cyclotron absorption line was fitted with the Lorentz profile
\citep{mih90}
$$\exp\left(\frac{-\tau_\mathrm{cycl}(E/E_\mathrm{cycl})^2\sigma_\mathrm{cycl}^2}
{(E-E_\mathrm{cycl})^2+\sigma_\mathrm{cycl}^2}\right),$$ where
$E_\mathrm{cycl}$, $\sigma_\mathrm{cycl}$, and $\tau_\mathrm{cycl}$ are the
line central energy, width, and depth, respectively. The fundamental
cyclotron line (below we'll call it ''the cyclotron line'' for a simplisity)
is clearly detected in all observations, whereas the second-harmonic line
was confidently observed not in all data sets. In some cases, inclusion of
the third harmonic in the model affected the parameters of the
second-harmonic line. The third harmonic itself was registered and its
parameters can be more or less reliably retrieved only for several brightest
states (see Paper~I for details). We examined possible effects of including
the second harmonic in the fit on the parameters of the cyclotron line and
the continuum. In these tests, the central energy of the second-harmonic
line was either a free parameter or fixed equal to two times the central
energy of the cyclotron line. It was found that the cyclotron line and
continuum parameters do not depend within errors on the higher harmonics.

%=================================================================
%=================================================================
\begin{figure} \includegraphics[width=\columnwidth,
clip]{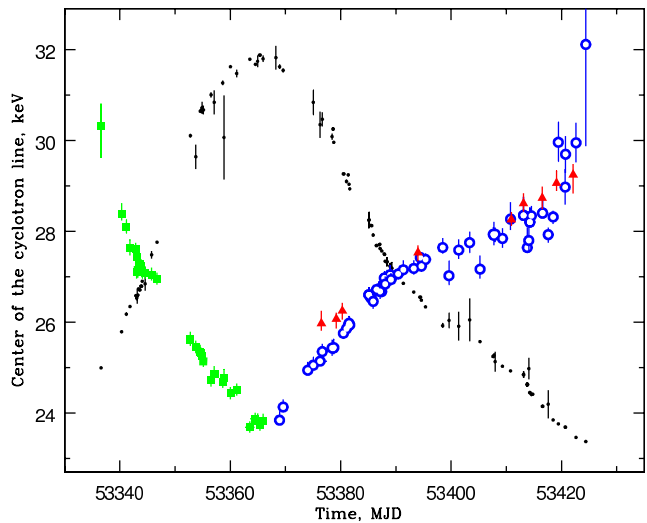} \caption{Temporal evolution of the
cyclotron-line energy during the V\,0332+53 outburst. The source bolometric
luminosity is shown with the small black circles. Other symbols are the same
as in Fig.\ref{lcurve}.}\label{cycvst} \end{figure}
%=================================================================
\section{Results}

This section presents the results of spectral analysis during almost
entire outburst of V\,0332+53. The analysis is focused on the source
spectral parameters as functions of the luminosity and time.

V\,0332+53 is one of two X-ray pulsars whose spectra exhibit at
least three cyclotron-harmonic absorption line
features~\citep{cob05,kre05}. In general, this source has a fairly
complex, multi-component spectrum. Therefore, we applied the
above-described spectral model including a power law with an
exponential cutoff, up to three absorption lines depending on the
source brightness and ''detectability,'' and the 6.4-keV iron
emission line which was clearly observed in all states of the
source. Since the iron line was narrow and the energy resolution of
the PCA spectrometer at $6\!-\!7$~keV was limited ($\sim\!18$\%),
the actual line width cannot be accurately determined. Thus, a fixed
value of $0.1$~keV was used in the XSPEC data analysis.

Variations in the main spectral parameters with the time and the
luminosity are described in detail
below.\vspace*{-\baselineskip}

\subsection{Cyclotron Energy Versus Luminosity}

It was mentioned above that variations in the cyclotron-line energy and their
possible dependence on the luminosity were discovered and reported by
\citet{mih98} for several bright X-ray pulsars. The
cyclotron-line\,energy\,--\,source\,luminosity relation was studied in
detail for the first time by \citet{tsy06} for V\,0332+53 in a wide
luminosity range. Using INTEGRAL and RXTE data obtained during the fading
phase of the 2004--2005 outburst, they discovered the strong negative
correlation between the cyclotron line energy and the luminosity
which can be described fairly well by a linear function.

\begin{figure}
\includegraphics[width=\columnwidth, clip]{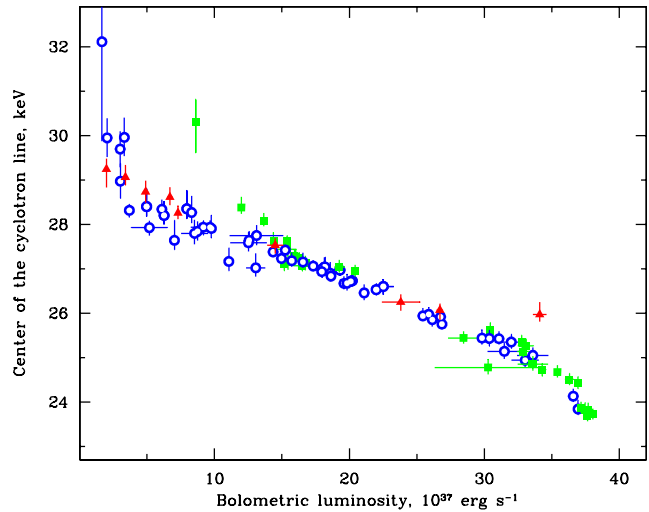}
\caption{The cyclotron-line energy versus the
luminosity $L_{37}$.  Symbols are
the same as in Fig.\ref{lcurve}.}\label{cycvslx}
\end{figure}

Including both the fading and brightening phases of the outburst into an
analysis makes it possible to study the possible difference in the
cyclotron-line behavior during these outburst stages to verify the
hypothesis on the hysteretic dependence of the source parameters on its
luminosity. Figure~\ref{cycvst} shows the temporal evolution of the
cyclotron-line energy during the entire outburst.  Distinct asymmetry of
this time profile can obviously be attributed to the asymmetry of the
outburst light curve. The existence of hysteresis the cyclotron-line energy
as a function of the source was checked by constructing this dependence for
the entire outburst (see Fig.~\ref{cycvslx}). The resulting plot
unambiguously rules out the hysteresis hypothesis. This is valid for 
luminosity above $\sim1.5\times10^{38}$ erg s$^{-1}$ (the
lower-luminosity data are absent in the brightening phase). Hence, the
physics of accretion 
column and its emission is essentially same at both these stages.

%=================================================================
%=================================================================

A formal linear fit of this dependence (all available measurements have been
used),
$E_\mathrm{cycl,1}\mathrm{[keV]}\!\simeq\!-\!0.143(\pm0.002)L_{37}\!+\!29.56(\pm0.03)$
(here, $L_{37}$ is the luminosity in the units of $10^{37}$~erg s$^{-1}$,
uncertainties correspond to a 90\% confidence level), is practically the
same as the one obtained in Paper I for only the fading phase of the
outburst and even takes into account the above-mentioned luminosity
corrections for several highest-count-rate observations. Note, however, that
allowance for these corrections makes the behavior of the cyclotron energy
at very high luminosities $>\!3.5\!\times\!10^{38}$ erg s$^{-1}$ slightly different
from this fit: the ''rate'' of the cyclotron-energy variations is
higher. Such a difference (if any) may be connected with possible changes in
the accretion-flow structure and emission at very high luminosities then the
source can transfer into other accretion regime. On the other hand, poor
statistics impedes finalizing conclusions on the behavior of the
cyclotron-line energy at low luminosities below $L_{37}\!\simeq\!5$. Some
evidence is seen in Fig.~\ref{cycvslx} for ''steepening'' of the
cyclotron\,energy\,--\,luminosity slope below
$\simeq\!4\!\times\!10^{37}$~erg s$^{-1}$, but its significance is not enough for
the final conclusion. Note, that the behavior of the cyclotron energy at low
luminosities is very interesting from the viewpoint of determining the
so-called critical luminosity at which the shock wave at the top of the
accretion column disappears and/or detecting the same reverse dependence (if
any) of the cyclotron-line energy with the luminosity as was observed for
Her X-1~\citep{sta07}.

Meanwhile, the lowest observed value of the cyclotron energy in the
V\,0332+53 spectrum provides for an estimate for the magnetic field
on the neutron-star surface. Unfortunately, the lowest-luminosity
data point in Fig.\,\ref{cycvslx} has fairly large uncertainty.
Hence, it seems more expedient to use the result from the linear
fit, that is 29.56 keV. This yields
$$B_\mathrm{NS}\!=\!(1+z)\!\times\!\frac{29.56}{11.6}\simeq3.1\times10^{12}~\mathrm{G},$$
where $z$ is the gravitational redshift near the surface of a
neutron star with a radius of 10 km and a mass of $1.4M_\odot$.

%=================================================================
\begin{figure}
\includegraphics[width=\columnwidth,bb=18 270 560 700, clip]{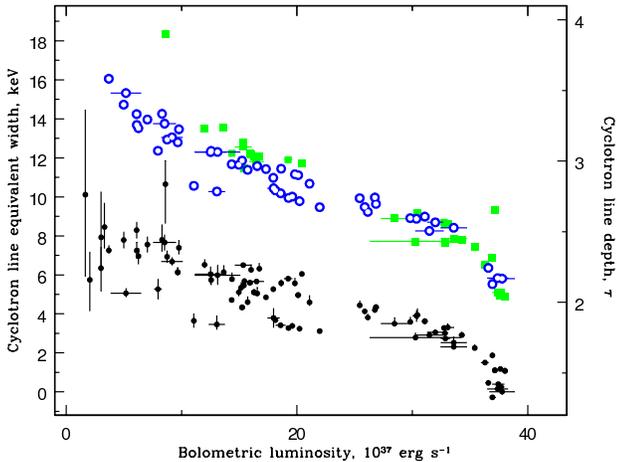} 

\caption{Dependence of the equivalent width $\mathrm{EW}_\mathrm{cycl}$
(open circles and filled squares) and the depth $\tau_\mathrm{cycl}$ (filled
circles) of the cyclotron line in V\,0332+53 spectra on the pulsar
luminosity $L_{37}$. Symbols are the same as in
Fig.\ref{lcurve}.}\label{eqwcycvslx}

\end{figure}
%=================================================================

\subsection{Equivalent Width of the Cyclotron Line}

It is well known from the classical stellar spectral analysis that the
equivalent width of a spectral line is a very important quantity for
retrieving physical parameters and conditions in the line-forming
region. Hence, in addition to such cyclotron-line model parameters as
$\sigma_\mathrm{cycl}$ and $\tau_\mathrm{cycl}$, it seems expedient to have
an observable quantity playing the role of cyclotron-line equivalent width
in X-ray astrophysics and quantifying the total number of photons withdrawn
from the continuum spectrum of a source due to resonance scattering in the
cyclotron line(s). The main problem in calculating the cyclotron-line
equivalent width is its strong dependence on the adopted continuum model to
be multiplied by the line models typically used to fit cyclotron-harmonic
lines. We used the following simple technique to calculate the
cyclotron-line equivalent width $\mathrm{EW}_\mathrm{cycl}$: (1)~All energy
channels affected by the cyclotron line (typically $18\!-\!40$~keV) are
excluded from the source spectrum; (2)~The remaining energy bins are fitted
as a spectrum of the continuum; (3)~The energy bins excluded at step~1 are
added again to the spectrum, their deviations from the continuum fit
obtained at step~2 is calculated, and, upon normalization to the continuum
spectrum, are integrated over energy; (4)~The resulting energy-integrated
deviation of the normalized cyclotron-line energy bins from the continuum
fit is adopted as a measure for the equivalent width
$\mathrm{EW}_\mathrm{cycl}$ of the cyclotron line\footnote{We should note,
that the spectrum of V0332+53 contains a few harmonics of the cyclotron line
and it is practically impossible to avoid fully the influence of absorption
features to the continuum component. Therefore, obtained values of the
equivalent width should be treated only for the demonstration of the
qualitative behaviour.  Such an analysis was carried out for sufficiently
bright observations.}.

It was found that the quantity $\mathrm{EW}_\mathrm{cycl}$ calculated in
such a way is not constant, but varies with the cyclotron-line energy
changes during the outburst. The $\mathrm{EW}_\mathrm{cycl}$ dependence on
the luminosity is shown in Fig.~\ref{eqwcycvslx}. As in Fig.~\ref{cycvslx},
different symbols such as filled squares and open circles correspond to the
brightening and fading phases of the outburst, respectively. The
cyclotron-line equivalent width decreases almost linearly with the
luminosity repeating the behavior of the cyclotron-line energy itself,
including the increase of the "rate" of the equivalent width changes at high
luminosities. But it is nessesary to note, that the slope of this formal
linear relation
$\mathrm{EW}_\mathrm{cycl}\mathrm{[keV]}\!\propto\!-\!0.25L_{37}$ is
different from that in Fig.3. It is also worthy to note the similar
behaviour of the cyclotron line optical depth $\tau_\mathrm{cycl}$ (see
filled circles in Fig.~\ref{eqwcycvslx}), that can be understood from the
revealed linear correlation between $\tau_\mathrm{cycl}$ and
$\mathrm{EW}_\mathrm{cycl}$.

%=================================================================
\begin{figure}
\includegraphics[width=\columnwidth,clip]{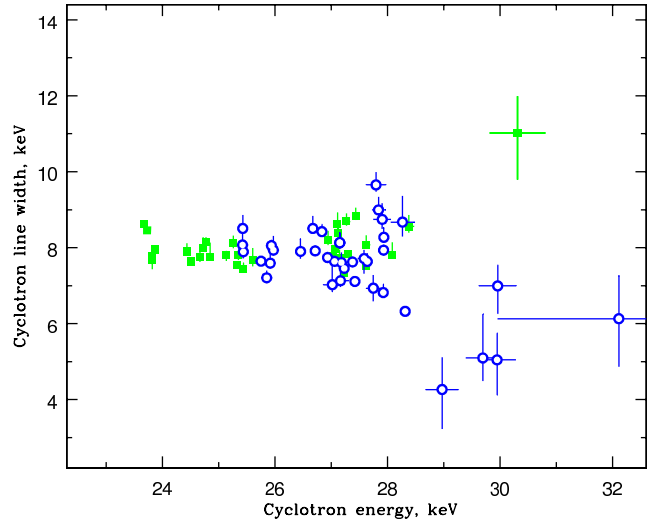}
\caption{Cyclotron-line width
$\sigma_\mathrm{cycl}$ versus cyclotron energy $E_\mathrm{cycl}$ for
all observed luminosities.  Symbols are
the same as in Fig.\ref{lcurve}.} \label{sigma_ecyc}
\end{figure}
%=================================================================

The correlation between the cyclotron-line energy $E_\mathrm{cycl}$ and its
width $\sigma_\mathrm{cycl}$ has been noted earlier in several works
\citep{hei99,dal00,cob02,sta03} for a number of X-ray pulsars with different
cyclotron line energies and viewing angles. The last one can play an
important diagnostic role. In particular, it was used by \citet{sta03} to
explain nondetection of cyclotron lines in certain objects.

Note, that the relation between $\sigma_\mathrm{cycl}$ and $E_\mathrm{cycl}$
stems from both the Doppler broadening due to thermal motion of electrons
along the magnetic field in the line-forming region and the natural
cyclotron-line broadening giving rise to broad Lorentz-shaped wings of the
line. The observed Doppler width of the cyclotron line
is~\citep{tru77,mes92}
$$\sigma_\mathrm{cycl,D}^\mathrm{obs}=E_\mathrm{cycl}(2\kappa
T_\mathrm{e}/mc^2)^{1/2}\langle\cos\alpha_\mathrm{obs}\rangle,$$
where $\langle ...\rangle$ denotes averaging over the duration of
one observation which is typically about a few kiloseconds, i.e.,
much longer than the X-ray pulsar period,
$$\alpha_\mathrm{obs}=\arccos[\cos\Theta_\mathrm{obs}\cos\Theta_*+\sin\Theta_\mathrm{obs}
\sin\Theta_*\cos\Phi_*(t)]$$ is the angle between the magnetic field
in the line-forming region and the observer line of sight,
$\Theta_\mathrm{obs}$ is the angle between the neutron-star spin
axis and the line of sight, $\Theta_*$ is the angle between the
neutron-star spin and magnetic axes, $\Phi_*$ is the angle between
the neutron-star magnetic axis and the plane comprising the
neutron-star spin axis and the line of sight, and $T_\mathrm{e}$ is
the electron temperature. In the absence of neutron-star precession
effects, $\Theta_\mathrm{obs}\!=\!\mathrm{const}$ over the binary
orbit. Hence,
$\langle\cos\alpha_\mathrm{obs}\rangle\!=\!\cos\Theta_\mathrm{obs}
\cos\Theta_*\!=\!\mathrm{const}$ and
$\sigma_\mathrm{cycl,D}^\mathrm{obs}/E_\mathrm{cycl}\!=\!(2\kappa
T_\mathrm{e}/mc^2)^{1/2}\cos\Theta_\mathrm{obs} \cos\Theta_*$. Note,
however, that the applied fit of the cyclotron line implies that the
natural broadening implying the Lorentz profile of the cyclotron
line is assumed dominant. In this case, the observed relative
cyclotron-line width
$\sigma_\mathrm{cycl,L}^\mathrm{obs}/E_\mathrm{cycl}\!=\!(4e^2/3\hbar
c)(E_\mathrm{cycl}/mc^2)\!\propto\!E_\mathrm{cycl}$ is independent
of the system geometry and the electron temperature in the source.
In addition, the ratio $\sigma_\mathrm{cycl}/E_\mathrm{cycl}$ can
also be contributed by other factors, e.g., nonuniformity of the
magnetic field over the line-forming region.

We examined the processed data for the presence of linear correlation
between $\sigma_\mathrm{cycl}$ and $E_\mathrm{cycl}$.  The cyclotron line
width as a function of its energy for V\,0332+53 during its 2004--2005
outburst is shown in Fig.~\ref{sigma_ecyc}. No pronounced linear increase in
$\sigma_\mathrm{cycl}$ with increasing $E_\mathrm{cycl}$ is seen. The
typical width of the cyclotron line is about $8$~keV. The fairly strong
scatter of the data points for large cyclotron energies is probably caused
by spectrum-statistics worsening at the low luminosities. This is confirmed
indirectly by an increase in the error-bar sizes. An additional factor which
affect to the observable broadening of the cyclotron line can be the finite
energy resolution of the HEXTE spectrometer (FWHM is about 5.5~keV at
30~keV).

Finally note that according to the theory of cyclotron-radiation transfer
under neutron-star conditions (see, e.g., \citep{ZhS93}), the optical depth
at the cyclotron line varies only slightly with $\alpha_\mathrm{obs}$ and
can hardly exclude the formation of a resonance scattering feature at fairly
large $\alpha_\mathrm{obs}$.  Hence, the cyclotron line nondetection should
rather be related to specific viewing conditions of the line-forming region
such as its small visible area or screening of this region by the neutron
star and/or accretion disc/stream.

%=================================================================
\begin{figure}
\vbox{
\includegraphics[width=\columnwidth,bb=86 310 505 690,clip]
{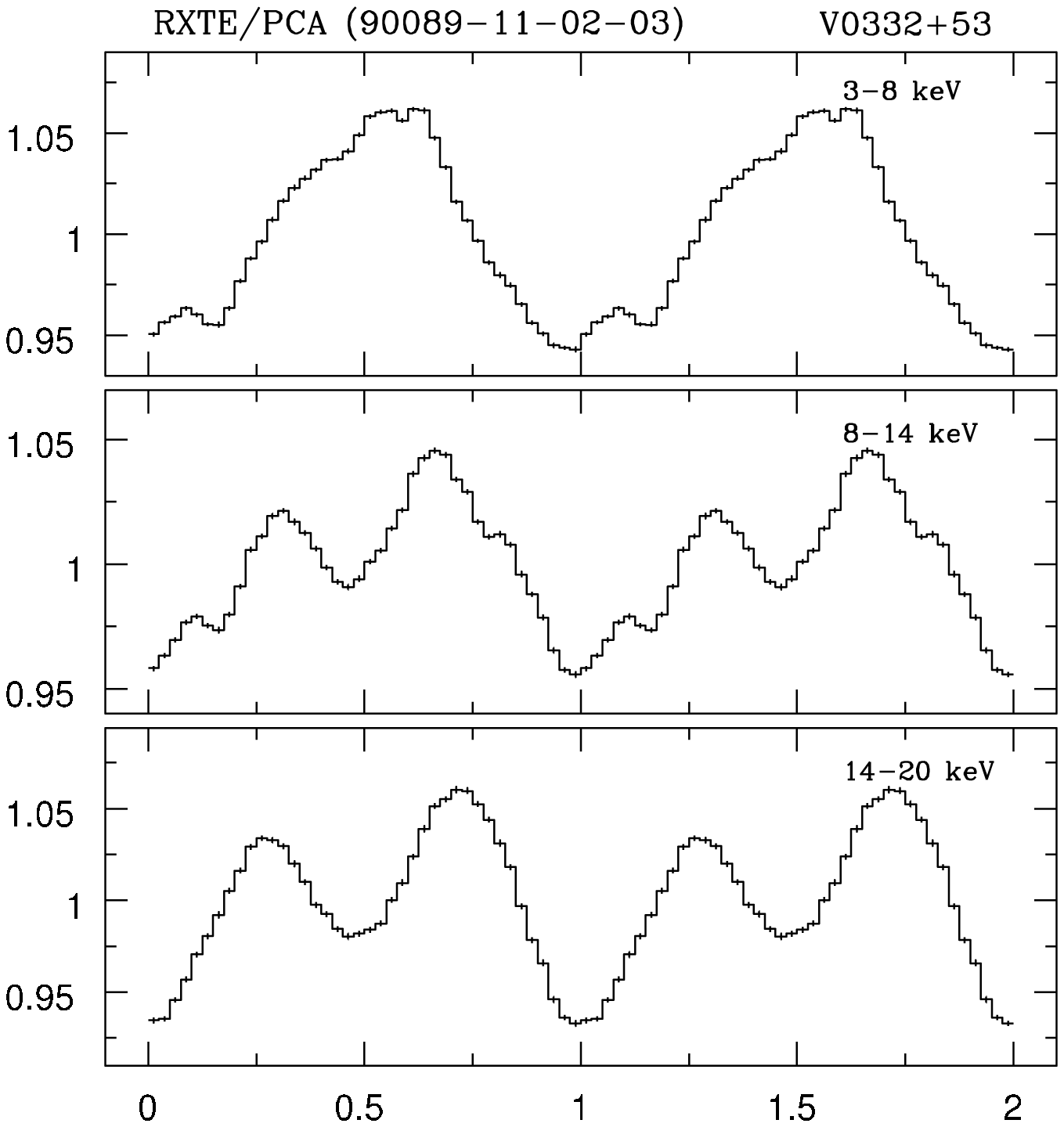}
\includegraphics[width=\columnwidth,bb=86 144 505 690,clip]
{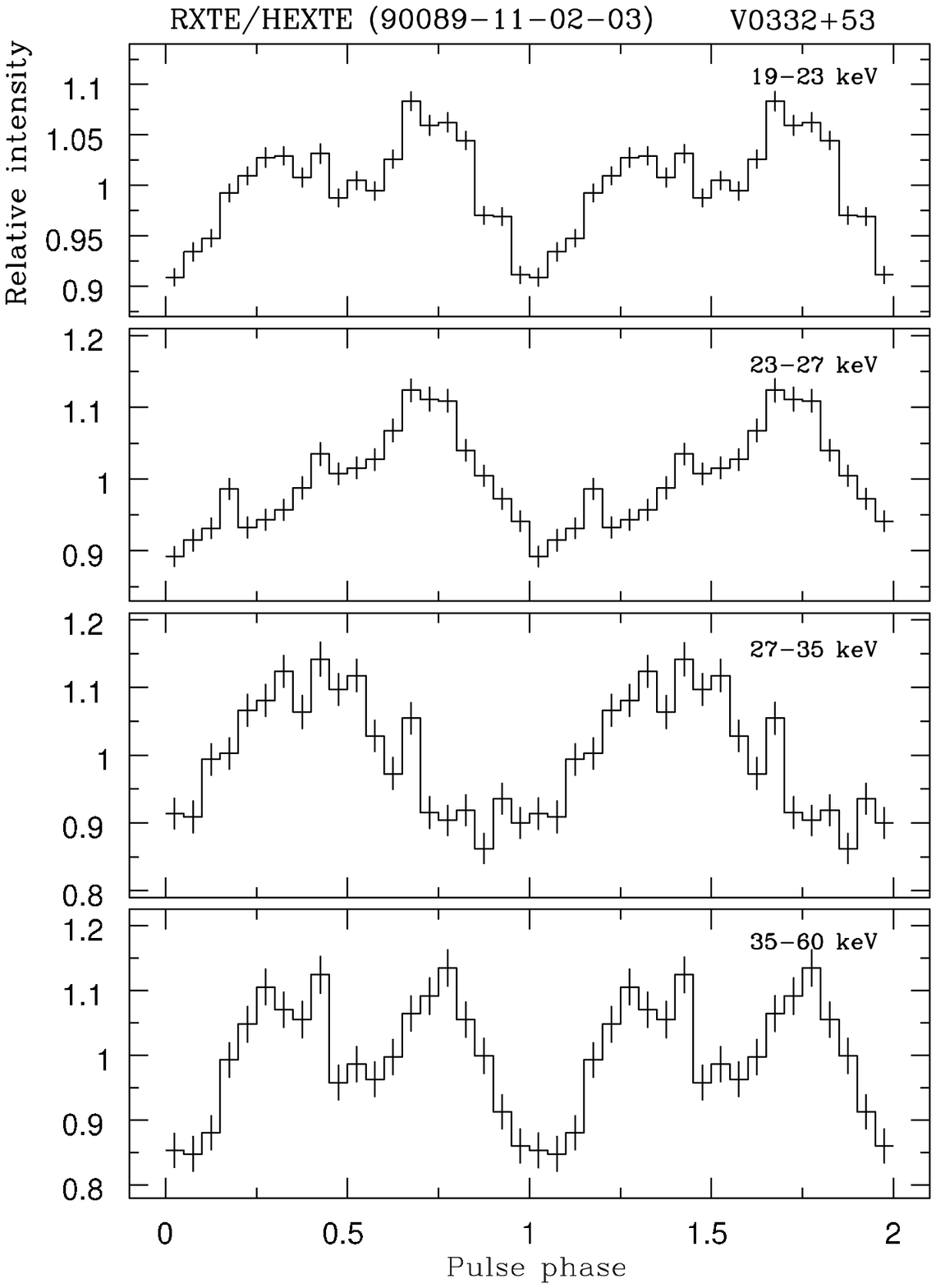} }

\caption{Pulse profiles of V\,0332+53 for the RXTE observation
90089-11-02-03 during the brightening phase of the 2004--2005
outburst for the luminosity $L_{37}\!\simeq\!16$ and the
cyclotron-line energy
$E_\mathrm{cycl}\simeq\!27.2$~keV}\label{pp_2d}
\end{figure}
%=================================================================

\subsection{Pulse-Profile Variations}

It is well known that pulse profiles of X-ray pulsars are strongly
dependent on the photon energy and luminosity. This is
especially prominent in soft energy bands strongly affected by the
intrinsic absorption in the binary.
As the photon energy increases toward the hard X-ray
range, the observed emission becomes progressively less sensitive to
the poorly known environment of the accreting neutron star and
carries more and more imprints of actual physical and geometrical
parameters in hot polar-cap regions on the neutron star where this
radiation is generated. Although these regions are believed to be
much simpler in structure and physics, strong changes of pulse
profiles are observed in their hard X-ray emission, as well. In
particular, this was demonstrated by \citet{lut09} who used INTEGRAL
data to reconstruct uniformly the pulse profiles of ten bright X-ray
pulsars in hard X-rays for different luminosities and study their
variability.

This section is aimed at answering the question on the absence or
presence of pulse-profile hysteretic behavior during the brightening
and fading phases of the outburst from V\,0332+53. It was shown in
the previous sections that the source spectrum and its parameters
are practically the same at the outburst stages with the same
luminosity. Taking into account the similarity of pulse profiles at
the same luminosities for the brightening and fading phases of an
outburst from 4U\,0115+63~\citep{tsy07} we can expect the similar
behavior for V\,0332+53.

According to Paper~I, the main mystery of the considered source is the
drastic change of its pulse profile near the cyclotron line. In particular,
it was shown that the pulse profile in the high-luminosity state
($L_{37}\!\sim\!35$) has a sinusoidal double-peaked shape at all
energies. The relative intensities of the pulses are slightly
different. Very small variations in the positions of the peaks depending on
the photon energy range were observed. As the pulsar luminosity decreases,
the pulse profile changes significantly. In the soft energy bands (at photon
energies less then $\simeq\!8$~keV) it becomes nearly single-peaked. One the
intra-peak gap disappears and the double-peaked profile transforms into an
asymmetrical single-peaked one at energies near the cyclotron line. Then the
pulse profile becomes double-peaked again with the further increase in the
photon energy.

Here, all the data covering the outburst of 2004--2005, including
the ID~90427 used in Paper I, are reanalyzed uniformly and a
detailed comparison of pulse profiles corresponding to the same
luminosities during the outburst brightening and fading phases is
made. It is found that they are similar to each other (minor
differences may only exist in the softest energy band $<\!8$~keV)
and their behavior with the luminosity identical to the
described above. Note, that for several highest-luminosity
observations it was impossible to produce the pulse profiles using
PCA data due to the absence of data modes with sufficient energy and
time resolution below 20~keV.

For illustrative purposes the pulse profiles in different energy bands for
the observation 90089-11-02-03 (the luminosity was
$1.6\times10^{38}$~erg s$^{-1}$ at the brightening phase) are shown in Fig.\ref{pp_2d}. The PCA channels
correspond to $3\!-\!8$, $8\!-\!14$ and $14\!-\!20$~keV to demonstrate
changes at soft photon energies and the transition from single- to
double-peak shape. Following Paper I, the hard energy channels (HEXTE data)
were chosen to divide the cyclotron line in equal parts.

%=================================================================
\begin{figure} \includegraphics[width=\columnwidth,bb=40 300 520 730,clip]
{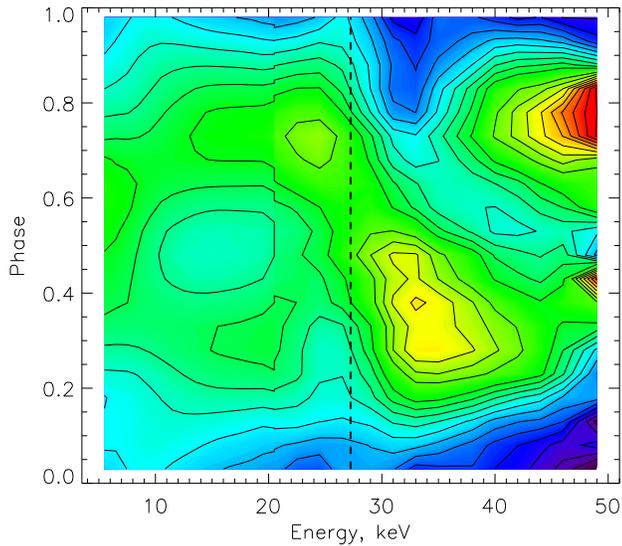}

\caption{Two-dimensional distribution of V\,0332+53 pulse-profile
intensities (normalized to unity) over the photon energy and the
pulse phase for the RXTE observation 90089-11-02-03 during the
brightening phase of the 2004--2005 outburst for the luminosity
$L_{37}\!\simeq\!16$. Position of the cyclotron line center is shown
by the dashed line. Red colour corresponds to higher and blue to 
lower relative intensities. Solid lines represent levels of
equal intensity: 0.20, 0.24, 0.28, ..., 0.96, 1.00. 
}\label{pp3d}
\end{figure}
%=================================================================

It is seen that relatively far from the cyclotron energy equal to
$\simeq\!27.2$~keV for this observation, the pulse profile is
double-peaked with a domination of a second peak in the photon
energy range $19\!-\!23$~keV. In the left (''soft'') wing of the
cyclotron line the profile is become single-peaked with its center
at the pulse phase of $\sim 0.65$. In the right (''hard'') wing of
the cyclotron line the profile is also single-peaked, but the
position of its maximum is shifted to the pulse phase $\sim\!0.3$.
At the higher energies the pulse profile becomes double-peaked
again. Note, that the observed shifting of the peak over the phase
near the cyclotron line energy is occurred gradually with increasing
photon energy. These features can be seen clearly in Fig.\ref{pp3d}
plotted using the technique described in detail in Paper I.

Note that the observed pulse profile is contributed by three main
components: (i)~direct radiation coming from one or two under-the-shock hot
polar columns which produce(s) the continuum and hosts the
cyclotron-line-forming region being presumably the upper part of the column
adjacent to the shock, (ii)~forward-scattered radiation outgoing from the
side of the accretion stream feeding the ``front'' (directly visible) polar
column and (iii)~backward-scattered radiation outgoing from the side of the
accretion stream feeding the ``back'' (neutron-star shaded) polar
column. Quantitative pulse-profile modeling is a complicated problem
requiring good knowledge of the features of radiation beaming near the
cyclotron frequency \citep{gs73,pav85}. In particular, it was shown by
\citet{mes85} that the accretion column beam function at the cyclotron
energy is significantly different from the radiation beaming at other
energies.

\subsection{Pulsed Fraction Dependence on the Energy and Luminosity}

It was already observed that the pulsed fraction
$\mathrm{PF}=(I_\mathrm{max}-I_\mathrm{min})/(I_\mathrm{max}+I_\mathrm{min})$,
where $I_\mathrm{max}$ and $I_\mathrm{min}$ are maximum and minimum
intensities in the pulse profile of an X-ray pulsar, respectively, increases
with energy in the hard X-rays ($>\!20$~keV) for all studied sources (see
e.g. \citet{sta80,fro85} and a last review of \citet{lut09} and references
therein). However, this increase in not monotonic for some sources and shows
such local features as, e.g., maxima near the cyclotron lines. The pulsed
fraction as a function of the photon energy was studied in detail in a wide
energy band ($3\!-\!100$ keV) for the pulsar 4U\,0115+63 using the RXTE data
\citep{tsy07}. It was found in this paper that (i)~the pulsed fraction
increases with the photon energy and decreased with the luminosity and
(ii)~the local PF maxima (hump-like features) exist not only near the
cyclotron line, but also near the higher harmonics.  Note that \cite{fer09}
analyzed archival BeppoSAX data for 4U0115+63 and found a local decrease in
the pulsed fraction near the second cyclotron harmonic in the middle of the
fading phase.

%=================================================================
\begin{figure}
\includegraphics[width=\columnwidth,bb=20 270 515 700,clip]
{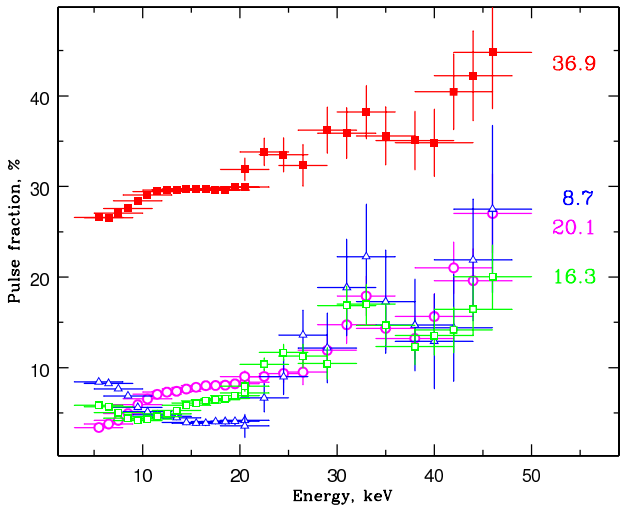}

\caption{Energy dependences of V\,0332+53 pulsed fraction during the
brightening and fading phases of the outburst. Different symbols
correspond to different source luminosities indicated to the right of
each plot in the units of $10^{37}$~erg s$^{-1}$. The green open and red
filled squares correspond to
$L_{37}\!\simeq\!16.3$ and 
$L_{37}\!\simeq\!36.9$, respectively,
during the brightening phase (RXTE observations 90089-11-02-03 and
90089-11-04-04, respectively). The magenta open circles and  blue
open triangles correspond to $L_{37}\!\simeq\!20.1$
and $L_{37}\!\simeq\!8.7$, respectively,
during the fading phase (RXTE observations 90014-01-02-00 and
90014-01-05-02, respectively).}\label{ppfr_pca_hex}
\vspace{\baselineskip}\end{figure}
%===================================================================

In this work, the V\,0332+53 pulsed fraction is analyzed in various
photon energy ranges and for different luminosity states using the
same complex approach that was employed by \citet{tsy07} for
4U\,0115+63.

Figure~\ref{ppfr_pca_hex} shows the energy dependences of the pulsed fraction
for observations with various pulsar luminosities obtained in a wide
photon-energy range for the brightening and fading phases of the
outburst. The corresponding luminosities in the units of $10^{37}$~erg
s$^{-1}$ are
indicated to the right of each plot. It is seen that the behavior of the
V\,0332+53 pulsed fraction at high energies ($>\!20$~keV) is typical of X-ray
pulsars (see, e.g., \citet{lut09}). In particular, the pulsed fraction
increases with the photon energy and the form of the increasing function
$\mathrm{PF}(E)$ depends on the luminosity, but changes only slightly
over a wide luminosity range and is roughly the same for similar
luminosities during the brightening and fading phases of the outburst.

On the contrary, these dependencies for soft energy bands are very
different. We should specially note the photon energies below
$12\!-\!15$~keV for which an increase in the pulsed fraction with
decreasing photon energy is observed for the first time. Such
behavior observed in a wide luminosity range is the most prominent
for luminosities $<\!1.5\!\times\!10^{38}$~erg s$^{-1}$. It is difficult to
explain it within the framework of known models since, in
particular, no significant absorption was observed in the source
spectrum during the entire outburst. Further detailed studies are
required to clarify this problem and we plan to perform such an
analysis in a separate paper. Here, we only note a steep rise in the
pulsed fraction from $\sim10$\% to $\sim30$\% with an increase in the
luminosity from $\sim2\times10^{38}$ to $\sim3.5\times10^{38}$~erg s$^{-1}$
and changes in the pulsed fraction behavior in the soft energy bands.

%===================================================================
\begin{figure}
\includegraphics[width=\columnwidth, clip]{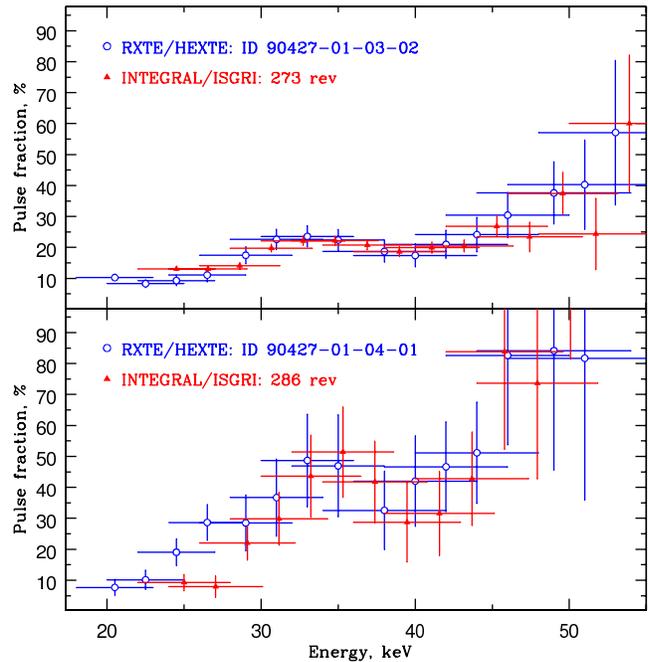}

\caption{Energy dependences of the V\,0332+53 pulsed fraction during
the high- (upper panel) and low-luminosity states (bottom panel).
The filled triangles represent INTEGRAL/ISGRI data during revolutions
273 (high state) and 286 (low state). The open circles are RXTE/HEXTE
data during observations 90427-01-03-02 (high state) and
90427-01-04-01 (low state). }\label{ppfr_hex_int}
\end{figure}
%=================================================================

It should also be noted that the V\,0332+53 pulsed fraction at soft
energies is small in comparison with values observed for other X-ray
pulsars. This fact was mentioned previously by \citet{mih07} who
attempted to explain this feature assuming that the angles
$\Theta_\mathrm{obs}$ and/or $\Theta_*$ are small. However, both
these assumptions can hardly be conformed with the observed high
values of the high-energy and high-luminosity pulsed fraction and the
observed energy and luminosity dependence of the pulsed fraction.

The most interesting feature at the high energies is a local pulsed fraction
maximum near the cyclotron-line energy. Such a hump-like feature is observed
for almost any luminosity during both brightening and fading phases of the
outburst. To rule out its possible artefact origin caused either by the
HEXTE data itself, we obtained the pulsed fraction dependence on the energy
using the INTEGRAL data (\citet{lut09}, see also
http://hea.iki.rssi.ru/integral/pulsars/index.php). Figure~\ref{ppfr_hex_int}
shows how the pulsed fraction changes with the energy for two values of the
luminosity. The RXTE/HEXTE and INTEGRAL/ISGRI results are shown by the
open circles and filled triangles, respectively. The upper and lower panel of
the figure correspond to the high and low luminosities $L_{37}\!\sim\!25$
and $L_{37}\!\sim\!5$, respectively.  It is clearly seen for both
luminosities that the pulsed fraction dependencies are practically the same
for the RXTE and INTEGRAL data and the ''hump''-like features are clearly
detected in both data sets. We analyzed all the available HEXTE data and
found that the ''hump'' energy ($E_\mathrm{hump}\!\simeq\!32\!-\!34$~keV) is
more or less stable over a wide luminosity range. This can imply that such a
pulsed fraction ''hump'' is possibly related to some intrinsic properties of
the neutron-star magnetosphere rather than to the local cyclotron energy
corresponding to the current luminosity. Finally, it is interesting to
note that the energy of this ''hump''-like feature is about of the maximum
measured value of the cyclotron energy for this source (see Section~3.1)
corresponding to the magnetic field at the surface of the neutron star.

It was shown by \citep{lut08,lut09} that the pulsed fraction in a
wide energy band usually decreases with increasing luminosity of an
X-ray pulsar. This fact was preliminarily explained within the
framework of a geometrical model in which the pulsed fraction is
determined by the luminosity-dependent visible areas of the
accretion columns.

%=================================================================
\begin{figure}
\includegraphics[width=\columnwidth, clip]
{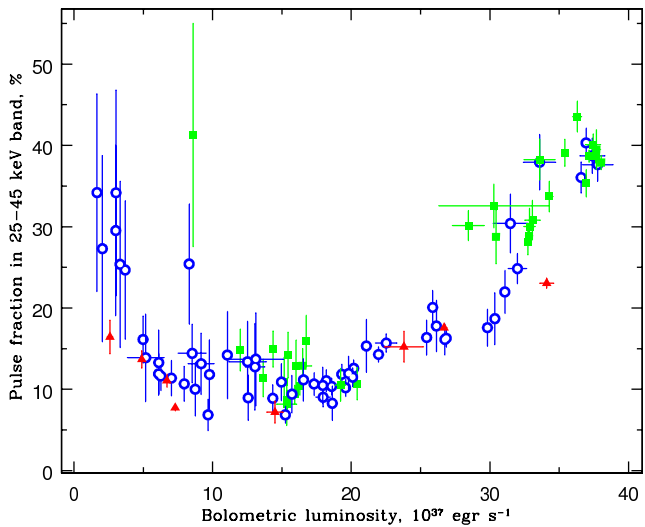} \caption{Pulsed fraction in the photon
energy range $25\!-\!45$~keV versus the V\,0332+53 bolometric
luminosity. As before, squares, circles and triangles of different
types represent the RXTE/HEXTE data covering the brightening and
fading phases of the outburst at rising and decay parts of the
outburst and the INTEGRAL/ISGRI data, respectively.
}\label{ppfrvslx}
\end{figure}
%=================================================================

The luminosity dependence of the pulsed fraction in the photon energy
range $25\!-\!45$~keV is shown in Fig.~\ref{ppfrvslx}. The observed
behavior does not fully conform to the predictions of the
above-mentioned model. Whereas the pulsed fraction decreases with
increasing luminosity below $\sim\!10^{38}$~erg s$^{-1}$ in accordance with
this model, its value remains almost constant at a level of
$\sim\!10$\% in the luminosity range
$(1\!-\!2)\!\times\!10^{38}$~erg s$^{-1}$ and rises almost linearly at
higher luminosities. Note that the luminosity dependence of the
pulsed fraction is the same for both brightening and fading phases of
the outburst.

\section{Summary}
The main results of the present work can be summarized as follows:

1.\ The brightening phase of the 2004--2005 outburst in the X-ray pulsar
V\,0332+53 is studied in detail for the first time, thereby completing an
analysis of the evolution of the accreting X-ray pulsar parameters over an entire
outburst.

2.\ It is shown that the equivalent width $\mathrm{EW}_\mathrm{cycl}$
demonstrates the strong positive linear correlation with the cyclotron-line
depth $\tau_\mathrm{cycl}$ and, similar the cyclotron-line energy
$E_\mathrm{cycl}$, decreases almost linearly with the bolometric luminosity
$L_\mathrm{X}$.

3.\ The hypothesis of hysteretic behavior of the observed radiation during
the brightening and fading phases of the outburst was checked and can be
ruled out. In particular, the source spectrum and its parameters and the
pulsed fraction behavior are practically the same at the outburst stages with
the same luminosity.

4.\ The presence of a local hump-like feature near the cyclotron-line energy
in the pulsed fraction as a function of the photon energy is confirmed. Its
position remains nearly constant at about $E_\mathrm{cycl,max}$.

5.\ An increase in the pulsed fraction with decreasing photon energy below
$12\!-\!15$~keV is observed for the first time in several observations.

6.\ The dependence of the pulsed fraction in the energy range
$25\!-\!45$~keV on the V\,0332+53 bolometric luminosity during the
brightening phase of the 2004--2005 outburst is found and compared with the
similar dependence for the fading phase of the same outburst. No hysteretic
behavior is observed. It was revealed for the first time that the pulse
fraction rises almost linearly at higher luminosities.

\section*{Acknowledgments}

We thank the referee for his useful comments and suggestions.  This work was
supported by the Russian Foundation for Basic Research (project
no.07-02-01051 and 08-08-13734), the Program ``Origin, Structure and
Evolution of the objects in the Universe'' (P07) by the Presidium of the
Russian Academy of Sciences and grant no.NSh-5579.2008.2 from the President
of Russia. We are grateful for the data to the High Energy Astrophysics
Science Archive Research Center Online Service provided by the NASA/Goddard
Space Flight Center.

\label{lastpage}


\begin{thebibliography}{99}

\bibitem[\protect\citeauthoryear{Basko \& Sunyaev}{1976}]{bs76}
Basko M.M., Sunyaev R.A., 1976, MNRAS, 175, 395

\bibitem[\protect\citeauthoryear{Bradt et al.}{1993}]{br93}
Bradt H.V., Rothschild R.E., Swank J.H., 1993, A\&AS, 97, 355

\bibitem[\protect\citeauthoryear{Coburn et al.}{2002}]{cob02}
Coburn W., Heindl W., Rothschild R., et al., 2002,
ApJ 580, 394

\bibitem[\protect\citeauthoryear{Coburn et al.}{2005}]{cob05}
Coburn W., Kretschman P., Kreykenbohm I., et al., 2005,
Astron. Telegram, 381, 1

\bibitem[\protect\citeauthoryear{Dal Fiume et al.}{2000}]{dal00}
Dal Fiume D., Orlandini M., Sordo S., et al., 2000, AdSpR, 25, 399

\bibitem[\protect\citeauthoryear{Frontera et al.}{1985}]{fro85}
Frontera F., Dal Fiume D., Morelli E.,  Spada G., 1985,  ApJ, 298, 585

\bibitem[\protect\citeauthoryear{Ferrigno et al.}{2009}]{fer09}
Ferrigno C., Becker P., Segreto A., et al., 2009,  A\&A, 498, 825

\bibitem[\protect\citeauthoryear{Filippova et al.}{2005}]{fil05} Filippova
C., Tsygankov S., Lutovinov A., Sunyaev R., 2005, Astron. Lett., 31, 729

\bibitem[\protect\citeauthoryear{Gnedin \& Sunyaev}{1973}]{gs73}
Gnedin Yu., Sunyaev R., 1973, A\&A, 25, 233

\bibitem[\protect\citeauthoryear{Heindl et al.}{1999}]{hei99}
Heindl et al. 1999, ApJ 521, L49

\bibitem[\protect\citeauthoryear{Kreykenbohm et al.}{2005}]{kre05}
Kreykenbohm I., Mowlavi N., Produit N., et al., 2005, A\&A, 433, L45

\bibitem[\protect\citeauthoryear{La Barbera et al.}{2005}]{lab05}
La Barbera A., Segreto A., Santangelo A., et al., 2005,  A\&A, 438, 617

\bibitem[\protect\citeauthoryear{Lutovinov \& Tsygankov}{2008}]{lut08}
Lutovinov A.A., Tsygankov S.S., 2008,  AIP Conference Proceedings, 1054, 191

\bibitem[\protect\citeauthoryear{Lutovinov \& Tsygankov}{2009}]{lut09}
Lutovinov A.A., Tsygankov S.S., 2009, Astron. Lett., 35, 433

\bibitem[\protect\citeauthoryear{Lyubarskii \& Sunyaev}{1988}]{ls88}
Lyubarskii Yu., Sunyaev R., 1988, Sov. Astron. Lett., 14, 390

\bibitem[\protect\citeauthoryear{Makishima et al.}{1990}]{mak90}
Makishima K., Mihara T., Ishida M., et al., 1990, ApJ, 365, L59

\bibitem[\protect\citeauthoryear{Meszaros}{1992}]{mes92}
Meszaros P., 1992, High-energy radiation from magnetized neutron
stars (univ. of Chicago Press)

\bibitem[\protect\citeauthoryear{Meszaros \& Nagel}{1985}]{mes85}
Meszaros P., Nagel W., 1985, ApJ, 299, 138

\bibitem[\protect\citeauthoryear{Mihara et al.}{1990}]{mih90}
Mihara T., Makishima K., Ohashi T. et al., 1990, Nature, 346, 250

\bibitem[\protect\citeauthoryear{Mihara et al.}{1998}]{mih98}
Mihara T., Makishima K., Nagase F. 1998, Adv. Space Res., 22, 987

\bibitem[\protect\citeauthoryear{Mihara et al.}{2007}]{mih07}
Mihara T., Terada Y., Nakajima M. et al., 2007, Progress of
Theoretical Physics, 169, 191

\bibitem[\protect\citeauthoryear{Negueruela et al.}{1999}]{neg99}
Negueruela I., Roche P., Fabregat J. et al., 1999, MNRAS, 307, 695

\bibitem[\protect\citeauthoryear{Pavlov et al.}{1985}]{pav85}
Pavlov G., Shibanov Yu., Silant'ev N., Nagel W., 1985, ApJ, 291, 170

\bibitem[\protect\citeauthoryear{Staubert et al.}{1980}]{sta80}
Staubert R., Kendziorra E., Pietsch W. et al., 1980, ApJ., 239, 1010

\bibitem[\protect\citeauthoryear{Staubert}{2003}]{sta03}
Staubert R., 2003, Chin. J. Astron. Astrophys., Vol 3, Suppl., 270

\bibitem[\protect\citeauthoryear{Staubert et al.}{2007}]{sta07}
Staubert R., Shakura N.I., Postnov K. et al., 2007, A\&A, 465, L25

\bibitem[\protect\citeauthoryear{Stella et al.}{1985}]{st85} Stella L.,
White N.E., Davelaar J., et al., 1985, ApJ, 288, L45

\bibitem[\protect\citeauthoryear{Swank et al.}{2004}]{sw04} Swank J.,
Remillard R., Smith E., 2004, Astron. Telegram, 349, 1

\bibitem[\protect\citeauthoryear{Tr\"umper et al.}{1977}]{tru77}
Tr\"umper J., Pietsch W., Reppin C. et al., 1977, Annals of the New
York Academy of Sciences, 302, 538

\bibitem[\protect\citeauthoryear{Tsygankov et al.}{2007}]{tsy07}
Tsygankov S.S., Lutovinov A.A., Churazov E.M., Sunyaev R.A., 2007, Astron. Lett., 33, 368

\bibitem[\protect\citeauthoryear{Tsygankov et al.}{2006}]{tsy06}
Tsygankov S.S., Lutovinov A.A., Churazov E.M., Sunyaev R.A., 2006, MNRAS, 371, 19

\bibitem[\protect\citeauthoryear{White et al.}{1983}]{wh83}
White N., Swank J., Holt S., 1983. ApJ, 270, 711

\bibitem[\protect\citeauthoryear{Zheleznyakov \&\ Serber}{1993}]{ZhS93}
Zheleznyakov V.V.\ and Serber A.V., 1993. Astron.\
Rep., 37, 507
\end{thebibliography}
\end{document}